\documentclass[a4paper]{article}

\usepackage{INTERSPEECH2019}
\usepackage{tikz}
\usepackage{color}
\usepackage{hyperref}
\usepackage{booktabs}
\newcommand*\circled[1]{\tikz[baseline=(char.base)]{
            \node[shape=circle,draw,inner sep=0.8pt] (char) {#1};}}
\newcommand{\squeezeup}{\vspace{-2.0mm}}

\title{ReMASC: Realistic Replay Attack Corpus for Voice Controlled Systems}
\name{Yuan Gong, Jian Yang, Jacob Huber, Mitchell MacKnight, Christian Poellabauer}
\address{Computer Science and Engineering, University of Notre Dame, IN 46556, USA}
\email{\{ygong1, jyang9, jhuber4, mmacknig, cpoellab\}@nd.edu}

\begin{document}

\maketitle
\begin{abstract}
This paper introduces a new database of voice recordings with the goal of supporting research on vulnerabilities and protection of voice-controlled systems (VCSs). In contrast to prior efforts, the proposed database contains both genuine voice commands and replayed recordings of such commands, collected in \textbf{realistic VCSs usage scenarios} and using \textbf{modern voice assistant development kits}. Specifically, the database contains recordings from four systems (each with a different microphone array) in a variety of environmental conditions with different forms of background noise and relative positions between speaker and device. To the best of our knowledge, this is the first publicly available database\footnote{Data available at \href{https://github.com/yuangongnd/remasc}{https://github.com/yuangongnd/remasc}} that has been specifically designed for the protection of state-of-the-art voice-controlled systems against various replay attacks in various conditions and environments.
\end{abstract}

\squeezeup
\section{Introduction}
\label{sec:intro}

Recently, an increasing number of voice-controlled systems (VCSs) have been introduced that rely on voice input as the primary user-machine interaction modality. For example, intelligent personal assistants such as Amazon Echo and Google Home allow users to control their smart home appliances, adjust thermostats, activate home security systems, purchase items online, initiate phone calls, and complete many other tasks with ease. VCSs also began to be used in vehicles to allow drivers to control their cars' navigation systems and other vehicle services. Despite their convenience, VCSs raise new security concerns due to their vulnerability to multiple types of spoofing attacks, such as replay attacks, self-triggered attacks~\cite{08diao2014your}, hidden voice commands~\cite{02vaidya2015cocaine,03carlini2016hidden}, and audio adversarial examples~\cite{06cisse2017houdini,05gong2017crafting,12carlini2018audio,gong2019realtime}. Such attacks pose significant threats, because they can easily be hidden and conducted remotely, and can be used to attack many systems simultaneously~\cite{d4gong2018overview}.

In order to defend against these attacks, the work presented in~\cite{d1blue2018hello} and~\cite{d3gong2018protecting} each proposes a defense strategy to protect VCSs by identifying the sound source of the received voice commands and rejecting those that are not from a human speaker, merely by analyzing the acoustic cues within the voice commands. This approach is based on the observation that legitimate voice commands should only come from human speakers rather than a playback device and that attacks such as self-triggered attacks, hidden voice commands, and audio adversarial examples rely on a playback device. That is, we are able to leverage the differences in the sound production mechanisms of humans and playback devices, which lead to differences in the frequencies and the directivity of the output voice signal. For example, in~\cite{d1blue2018hello}, the authors leverage the presence of significant low-frequency signals to distinguish electronic speakers from human speakers, while in~\cite{d3gong2018protecting}, the authors use a combination of features including fundamental frequency and Mel-frequency cepstral coefficients (MFCCs), and propose a data-driven approach.

The key idea in~\cite{d1blue2018hello, d3gong2018protecting} is actually an extension of the anti-spoofing technologies used for protecting automatic speaker verification (ASV) systems. Many prior efforts (such as presented in~\cite{jelil2017spoof,witkowski2017audio,02todisco2017constant,bakar2018replay,lavrentyeva2017audio,li2017study,cai2017countermeasures,wang2017feature,chen2017resnet,Jelil2018,kamble2018effectiveness,Yang2018}) have attempted to differentiate between original and replayed speech using the {\em RedDots Replayed} data set~\cite{kinnunen2017reddots}. While the VCS and ASV protection tasks look similar, they have some important differences, e.g., they have fundamental different user scenarios: ASV systems usually assume that the user speaks in a controlled environment and in close proximity to the systems, while modern VCSs usually support far-field speech recognition and are often used in a variety of environmental conditions indoors and outdoors. With increasing distance, the effects of environmental noise grow rapidly, which may impact the features the protection model relies on. In addition, modern VCSs usually feature microphone arrays, which can assist with sound source identification through directivity cues, while ASV systems usually have a single microphone only. We provide a more detailed discussion of these differences in Section~\ref{sec:2}. Therefore, since the RedDots Replayed data set has been recorded with short speaker-device distances, in indoor environments, and with devices that use a single microphone, it is not a suitable choice for research on VCS protection. Consequently, prior VCS protection research~\cite{d1blue2018hello,d3gong2018protecting} relied on self-collected non-public data for their experiments, where these data sets typically contain samples from a small number of subjects ($\le8$), limited environmental settings and number of voice commands ($\le8$), and only in single-microphone settings. The 2019 ASVspoof challenge~\cite{asvspoof2019} provides \emph{simulated} data for clear theoretical analysis of audio spoofing attacks in physical environments, but also leaves the {\em simulation-to-reality gap}. Therefore, in order to facilitate future research on the protection of VCSs, we present the {\bf ReMASC} (Realistic Replay Attack Microphone Array Speech Corpus) data set, which, compared to the RedDots Replayed data set and the data sets used in~\cite{d1blue2018hello,d3gong2018protecting}, has more {\em data variety} and is closer to realistic VCS usage scenarios and settings. Specifically, the data set contains recordings from 50 subjects of both genders and with different ages and accents. The recordings have been obtained in four different environments (two indoor, one outdoor, and one moving vehicle scenario) with varying types and levels of noise, and consisting of 132 voice commands. The distance between speaker and device varies from 0.5m to 6m, the data set for the indoor environments uses different placements of the VCS device, and different VCS devices (with different microphone configurations) are used.

\section{Comparing VCS and ASV Protection}
\label{sec:2}

While prior work has investigated approaches to differentiate between original and replayed voices, the focus has been on anti-spoofing of ASV systems. At first glance, VCSs and ASV systems appear similar, but there are important differences that prevent us from directly reusing data sets designed for ASV protection for research on VCS protection.

First, in ASV applications, the microphone is usually positioned close to the user (i.e., less than 0.5m). At such short distances, certain acoustic features can be used to identify the sound source of the speaker, e.g., in~\cite{104shiota2016voice}, the authors use the ``pop noise'' caused by breathing to identify a live speaker. Other efforts~\cite{Wickramasinghe2018,114korshunov2018use} do not explicitly use close-distance features, but the databases they use to develop their defense strategies were recorded at close distances~\cite{kinnunen2017reddots,Korshunov2016Overview,violatoBio}, and therefore, these approaches may also implicitly use close-distance features. In contrast, with the help of \emph{far-field speech recognition} techniques, modern VCSs can typically accept voice commands from rather long distances (i.e., several meters)~\cite{d3gong2018protecting}. At such distances, close-distance features cannot be used to distinguish between human speakers and recorded voice, e.g., the pop noise effect quickly disappears over larger distances, and the increasing effect of environmental noise may impact the features the protection model relies on. Further, modern VCSs usually allow the user to use it in a variety of environments, which also increases the protection challenge.

Second, modern ASV systems typically use a strict speaker verification model. Therefore, an attacker must either secretly record (e.g., via telephone or far-field microphones) or synthesize (e.g., using voice conversion or cutting and pasting) the victim's voice as a malicious voice command~\cite{15mukhopadhyay2015all}. In both cases, various cues, such as channel and background noise~\cite{100villalba2011detecting,119villalba2011preventing} or cutting-pasting traces~\cite{119villalba2011preventing}, will be left in the source recording used to replay (i.e., \circled{2} in Figure~\ref{fig:replayAttack}), which can be used to detect the attack. In contrast, for usability considerations, VCSs typically use less strict speaker verification, e.g., in~\cite{zhang2017dolphinattack}, the authors report that similar voices can activate Siri. In fact, speaker verification is not a mandatory or default setting of many VCSs such as Google Home or Amazon Alexa, while other VCSs do not even have a speaker verification function (e.g., Xiaomi MI AI, a smart home control device). 
This makes it easier for an attacker to obtain a clean source recording for replay, e.g., by recording the voice from a person with a similar pitch at a close distance or by synthesizing a similar voice or building an adversarial example. 
Therefore, a robust defense model for VCSs should 
focus on detecting differences in the playback phase.

Third, modern VCSs typically rely on microphone arrays, which allows them to perform far-field speech recognition, while ASV systems usually use a single microphone. For example, the Amazon Echo Dot has a 7-microphone array and Google Home Mini has a 2-microphone array. This can be an important characteristic for future research, i.e., a microphone array could be used to detect the directivity of the sound source or conduct noise canceling before spoof detection. However, existing data sets ignore this completely and only provide recordings using a single microphone.

\squeezeup
\section{ReMASC Data Collection}

\subsection{Definitions and Data Collection Strategy}

\begin{figure}[t]
  \centering
  \includegraphics[width=7.9cm]{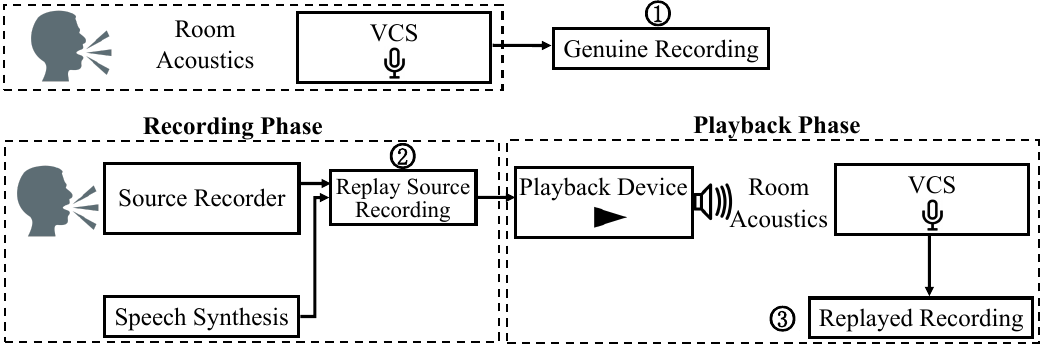}
  \squeezeup
  \caption{An illustration of legitimate usage of a VCS (upper figure) and a replay attack (lower figure).}
  \squeezeup
  \label{fig:replayAttack}
\end{figure}

\begin{figure}[t]
  \centering
  \includegraphics[width=6.5cm]{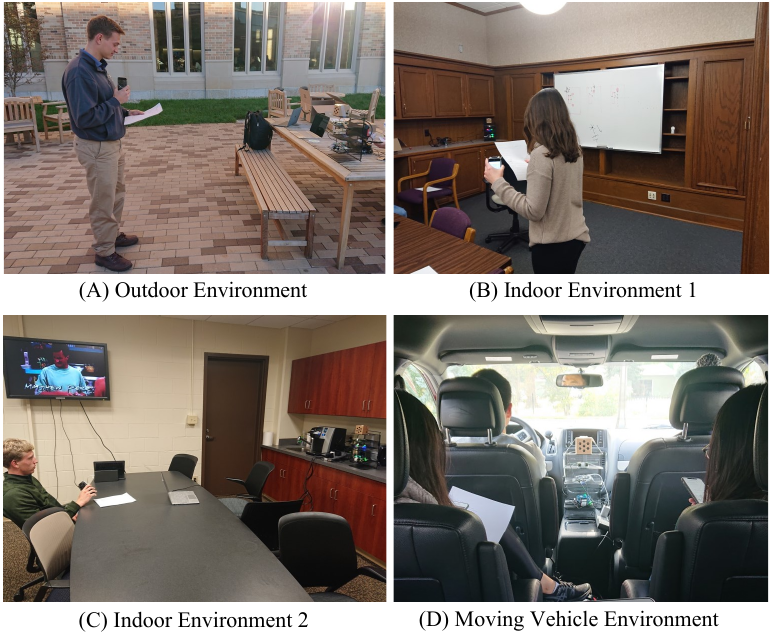}
  \caption{The recording environments and conditions.}
  \label{fig:env}
  \squeezeup \squeezeup
\end{figure}

A typical VCS replay attack is illustrated in the lower part of Figure~\ref{fig:replayAttack}. An attacker first needs to prepare a \textbf{replay source recording} (i.e., \circled{2} in Figure~\ref{fig:replayAttack}), which can be done by either recording a speaker (using a source recorder) or by performing speech synthesis. The attacker can then replay it using a replay device and the \textbf{replayed recording} (i.e., \circled{3} in Figure~\ref{fig:replayAttack}) is captured by the VCS device. In contrast, a legitimate usage scenario is illustrated in the upper part of Figure~\ref{fig:replayAttack}, where a \textbf{genuine recording} (i.e., \circled{1} in Figure~\ref{fig:replayAttack}) is directly captured by the VCS device. A defense task is then to build a model that is able to \textbf{distinguish genuine recordings from replayed recordings}. As shown in Figure~\ref{fig:env}, in our data collection, the subject holds the source recorder in the hand when speaking into the microphone arrays (which emulates the VCS device). When the subject speaks the voice command, both the source recorder and the microphone array record simultaneously. We define the recording captured by the microphone array as the \textbf{genuine recording}, and the recording captured by the source recorder as the \textbf{replay source recording}. Then, we play the replay source recording multiple times in different settings into the microphone array again, and refer to the recording captured by the microphone array as the \textbf{replayed recording}. The ReMASC data set provides all three types of recording. Note that different from previous work~\cite{kinnunen2017reddots}, we do not use genuine recordings as replayed source recordings because they may contain a high level of noise in our far-field setting. We also emulate situations where the attacker uses speech synthesis to generate replay source recordings (i.e., there is no genuine recording).

\squeezeup
\subsection{Text Materials and Recording Subjects}

A total of 132 voice commands are used as the recording text material. Among them, 31 commands are security sensitive and 49 commands are used in the vehicle. The command list contains 273 unique words, which provides reasonable phonetic diversity. We recruited 50 subjects (22 female and 28 male), where 36 subjects are English native speakers, 12 subjects are Chinese native speakers, and 2 subjects are Indian native speakers. The subjects' ages range from 18 to 36. Three subjects participated more than once, leading to a total of 55 sets (i.e., 47 subjects with one set and 3 with several sets of recordings). 

\begin{table}[t]
\scriptsize
\centering
\caption{Microphone array settings}
\squeezeup
\label{tab:setting}
\begin{tabular}{@{}cccc@{}}
\toprule
Device             & Sample Rate & Bit Depth & \#Channels \\ \midrule
Amlogic 113X1      & 16000       & 16       & 7          \\
Respeaker 4 Linear & 44100       & 16       & 4          \\
Respeaker V2       & 44100       & 32       & 6          \\
Google AIY         & 44100       & 16       & 2          \\ \bottomrule
\end{tabular}
\squeezeup \squeezeup \squeezeup
\end{table}

\begin{figure}[t]
  \centering
  \includegraphics[width=6.3cm]{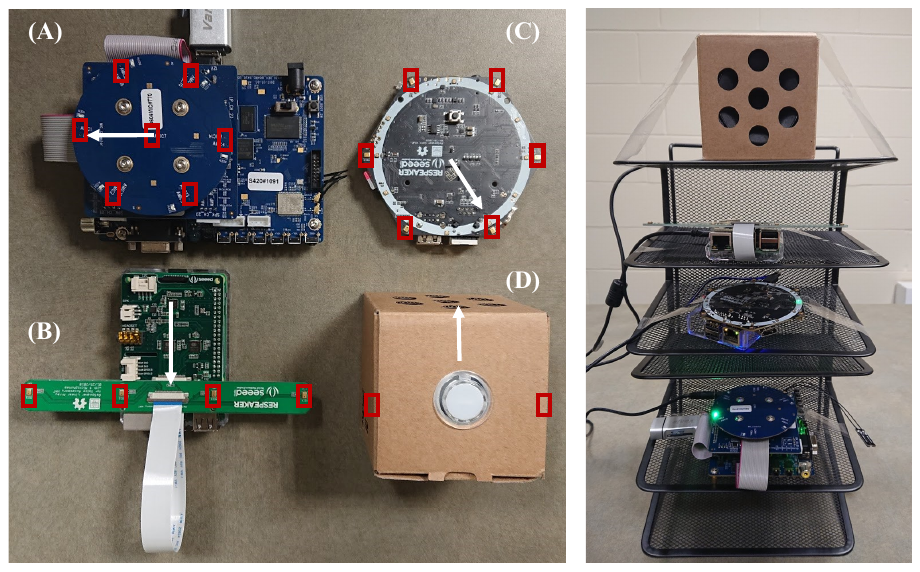}
  \caption{Microphone arrays used in the data collection (microphones are shown with the rectangles and a arrow indicates the direction of the microphone array during data collection).}
  \squeezeup
  \label{fig:VCS}
  \squeezeup \squeezeup
\end{figure}

\begin{figure}[h]
  \centering
  \includegraphics[width=7.5cm]{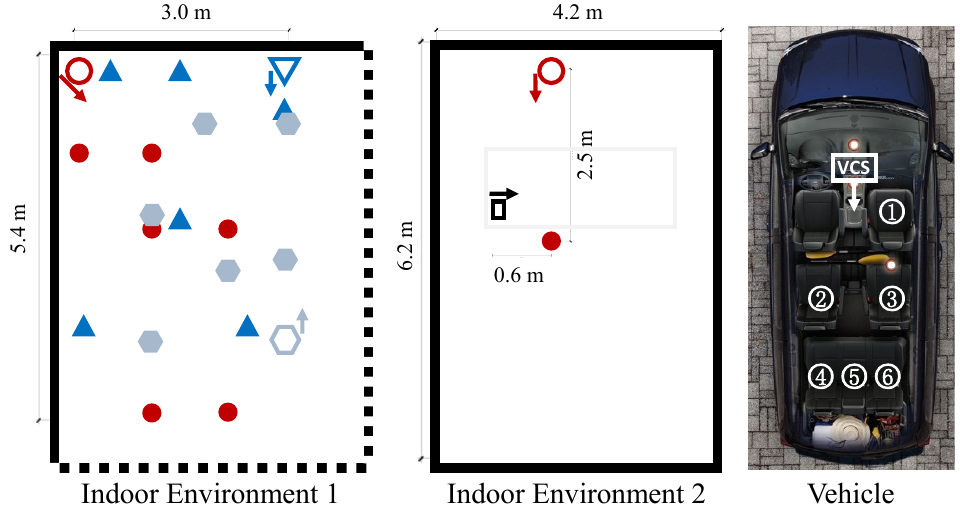}
  \caption{Illustration of device and speaker position settings. In indoor environment 1, each hollow symbol represents a microphone array placement and the direction it faces is indicated by an arrow. The corresponding solid symbols of the same shape represent a speaker position (for a total of 18 device placement - speaker position combinations, can be generalized to more combinations since the array is symmetric). In indoor environment 2, the hollow circle represents the microphone array, the square represents the speaker playing the background sound, and the solid circle represents the speaker. In the moving vehicle environment, the white square represents the microphone array placement and the direction it faces is indicated by the arrow; the circles represent the speaker positions.}
  \squeezeup
  \label{fig:position}
\end{figure}

\squeezeup
\subsection{Microphone Array Based Recorder}

Due to privacy concerns, off-the-shelf VCS products such as Amazon Echo or Google Home do not allow developers to access the raw audio. Therefore, we use the following VCS development kits in our work: A) Amlogic A113X1 (4-mic triangle or 6-mic circular array); B) Respeaker 4-mic linear array; C) Respeaker Core V2 (6-mic circular array); and D) Google AIY Voice Kit (2-mic linear array). As illustrated in Figure~\ref{fig:VCS}, in all experiments, we mount the four microphone arrays on a stand and for all recording devices, we use the Advanced Linux Sound Architecture (ALSA) to collect multi-channel waveform files. We use the highest possible recording quality for each kit (summarized in Table~\ref{tab:setting}). 
Practical VCSs might use lower sampling rates and bit depths to lower the computational and network transmission overheads.

\subsection{Source Recorder and Playback Devices}

\begin{figure}[t]
  \centering
  \includegraphics[width=4.7cm]{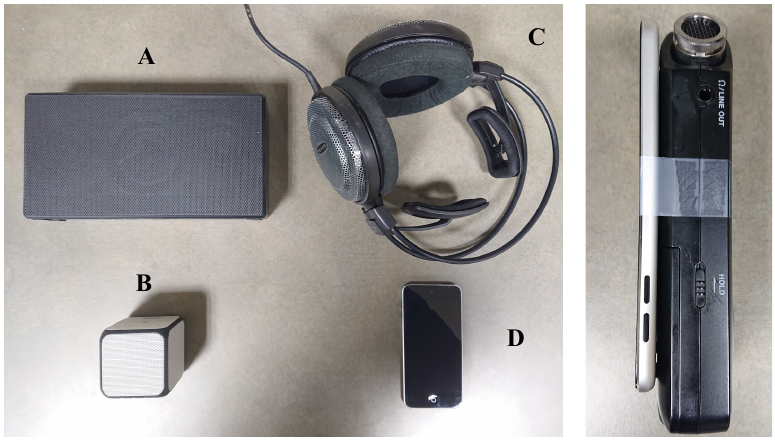}
  \caption{Playback device (left figure) and source recorder (right figure) used in the data collection.}
  \label{fig:playback}
  \squeezeup
  \squeezeup
  \squeezeup
\end{figure}

As discussed in Section~\ref{sec:2}, in the worst case, the attacker may have a high-quality replay source file. A robust defense model should still be able to detect the replay attack. To study if the source recorder affects the replay attack detection, we use a low-cost recorder, i.e., an iPod Touch (Gen5), and a professional recorder, i.e., a Tascam DR-05, together as the source recorder. As shown in Figure~\ref{fig:playback}, we tape the two recorders together and ask the subject to hold it at a close distance when they speak into the VCS (microphone array). The captured recording is then used as the replay source recording. Although the Tascam DR-05 is a professional high fidelity device, channel and background noise are still inevitable. Therefore, we also use Google Text-to-speech (TTS) to synthesize the voice commands as additional replay source recordings, which can then be considered as completely channel and background noise free. For diversity considerations, we use 26 different voice settings (13 male and 13 female) with two different synthesis technologies (standard and WaveNet) and three dialect settings (Australia, UK, and U.S.). As shown in Figure~\ref{fig:playback}, we use four common representative playback devices: A) Sony SRSX5, B) Sony SRSX11, C) Audio Technica ATH-AD700X headphone, and D) iPod Touch. Further, in the vehicle environment, we use the built-in vehicular audio system (of a Dodge Grand Caravan) as an additional playback device (i.e., connect an iPod to the car's audio system).

\squeezeup
\subsection{Recording Environment}

We performed the data collections in four environments:\\
{\bf Outdoor environment (Env-A):} To emulate uncontrolled noise in outdoor environments, we collected data on a student plaza with various background noises such as chatting, traffic, and wind. The speaker-recorder distance ranges from 0.5-1.5m.\\
{\bf Indoor environment 1 (Env-B):} Modern VCSs usually allow flexible device placement and speaker positions. To emulate this, we performed data collections in a quiet study room using three device placement settings: corner of the room, against the wall, and center of the room. For each device placement, the speaker spoke in six locations, forming 18 different position combinations (illustrated in Figure~\ref{fig:position}). \\
{\bf Indoor environment 2 (Env-C):} In realistic scenarios, VCSs will receive voice commands while some background sounds might be playing. In such situations, although there is an electronic device playing sounds, the VCS should not reject the user's voice command. This requires a defense model that is able to precisely detect the sound source of the voice command rather than that of the entire received audio. To emulate this situation, we collected data in a lounge where music players and TVs were running in the background (illustrated in Figure~\ref{fig:position}). Device and speaker positions were fixed.\\
{\bf Vehicle environment (Env-D):} To emulate a vehicle-based VCS, we collected data in a moving vehicle (Dodge Grand Caravan). As shown in Figure~\ref{fig:position}, the microphone array was placed at the center console and the subjects spoke while sitting in different seats (except the driver's seat for safety consideration). It is very common that the driver will make voice commands; therefore, about half of the data was collected from position \circled{1} in Figure~\ref{fig:position}. Each subject was asked to say 49 vehicle-related voice commands twice, once in a silent environment (parking lot when the engine is off) and once when the car is moving. The source recording obtained from a silent environment was then used for replay. We collected the data in various environments (e.g., campus, residential area, urban area, and highway), with speeds ranging from 3 to 40 miles per hour.

\squeezeup
\subsection{Replay Settings}
\squeezeup

For each replay source recording collected by each source recorder, we replayed it multiple times with different playback devices. In indoor environment 2 and the vehicle environment, the position of the playback device was identical to the subject's position. In the outdoor environment and indoor environment 1, we also replayed it in different positions. To keep the data collection effort reasonable, each replay source recording was replayed in 1 to 3 randomly selected replay settings, while the replay settings are uniformly distributed. All replay and genuine recordings were collected in the same environments with similar volume. Further, for each recording environment, we did our best to make everything in the environment identical for both genuine recordings and replay recordings. As shown in Table~\ref{tab:data_volume}, 9,240 genuine and 45,472 replayed recordings were collected (a recording captured by each microphone array is regarded as one recording regardless of the number of microphones). 

\squeezeup
\subsection{Data Availability}
\squeezeup

The ReMASC corpus is publicly available online for research purpose. Currently, two disjoint sets have been released. First, the {\fontfamily{pcr}\selectfont Quick Evaluation Set} consists of a small number ($\sim$2,000) of representative samples covering all recording conditions. This set can be used for quick evaluation of the performance of existing anti-spoofing models (e.g., models trained on RedDots Replayed data set) in the realistic settings of our data set. Second, the {\fontfamily{pcr}\selectfont Core Set} consists of $\sim$27,000 samples, which allows a user to build, validate, and evaluate the defense model as well as analyze the impact of factors such as the type of playback device and microphone. The rest of the data is reserved as an additional evaluation set for future defense model comparison and will be released in the future.

\squeezeup

\section{Experimentation and Conclusions}

We performed four baseline experiments and present the results in Table~\ref{tab:baseline}. For comparison with the RedDots Replayed data set, we use the first channel of each multi-channel audio and downsample it to 16KHz for the ReMASC data set. Since the purpose is to study the impact of the data set, we fix the classification algorithm for all experiments. Specifically, we use the official ASVspoof 2017 Challenge baseline CQCC-GMM model~\cite{Kinnunen2017} using constant Q cepstral coefficients (CQCC)~\cite{02todisco2017constant,todisco2016new} features and Gaussian mixture model (GMM) classifier and use exactly the same hyper-parameters. We further use conventional equal error rate (EER) as the metric.

\textbf{First}, we train the baseline CQCC-GMM model using the Reddot Replayed data set (training + development set), and then test it on the ReMASC data set. This is to evaluate if the defense model trained with data collected in partially controlled indoor environments with short speaker-microphone distances can be generalized to realistic VCS usage scenarios. The trained defense model (referred to as RedDots Pre-trained in Table~\ref{tab:baseline}) achieves 23.9\% EER (our reproduction, slightly differs from the result of 24.7\% reported in~\cite{Kinnunen2017}) on the evaluation set of the RedDots Replayed data set, but performs much worse on the ReMASC data set (note that the lower bound of EER is 50\%), indicating that the performance of the anti-spoofing model is sensitive to the environment and replay/recording settings, and may fail to work in (unseen) realistic scenarios. \textbf{Second}, we train the baseline CQCC-GMM model with both RedDots Replayed (training + development set) and entire ReMASC data set together, then test it on the RedDots Replayed evaluation set and achieve an EER of 20.2\%, which is 3.7\% lower than the EER achieved by the model trained with only the RedDots Replayed data set, indicating that training a defense model with additional data collected in various uncontrolled realistic conditions can also improve its performance in relatively controlled conditions. \textbf{Third}, we evaluate the defense performance when the target environment is unseen by the model using the ReMASC data set. Specifically, when testing in each target environment, we train the baseline CQCC-GMM model using data of three environments other than the target environment (referred to as Env-Independent in Table~\ref{tab:baseline}). We observe that the trained models perform noticeably better than the RedDots Pre-trained model (except Env-D), but are still unsatisfactory, especially when the speaker speaks from various distances to the microphone (e.g., Env-B) or the environment has complex noise (Env-C \& D). This indicates that the environment and recording scenario do have a large impact on the defense model. \textbf{Fourth}, we evaluate the defense performance when the target environment is seen by the model. For each environment, we split the ReMASC data set randomly into two disjoint and speaker-independent sets of roughly same size and then train the baseline CQCC-GMM defense model (referred to as Env-Dependent in Table~\ref{tab:baseline}) on one set and test on the other. We observe a remarkable improvement compared with RedDots Pre-trained and the Env-Independent model, indicating that the defense model can be significantly strengthened if it has knowledge about the target environment, even when the speaker is unknown. 

\begin{table}[t]
\centering
\footnotesize
\caption{Data volume of the ReMASC corpus (* indicates incomplete data due to recording device crashes).}
\squeezeup
\label{tab:data_volume}
\begin{tabular}{lccc}
\hline
Environment & \# Subjects & \# Genuine & \# Replayed\\ \hline
Outdoor     & 12          & 960                 & 6,900             \\
Indoor 1    & 23          & 2,760*              & 23,104            \\
Indoor 2    & 10          & 1,600               & 7,824             \\
Vehicle     & 10          & 3,920               & 7,644             \\
Total       & 55          & \textbf{9,240}      & \textbf{45,472}   \\ \hline
\end{tabular}
\squeezeup
\end{table}

\begin{table}[]
\footnotesize
\caption{Accuracy of baseline countermeasures (EER, \%) in various environments of the ReMASC data set.}
\squeezeup
\label{tab:baseline}
\centering
\begin{tabular}{@{}lccccc@{}}
\toprule
                   & Env-A & Env-B & Env-C & Env-D \\ \midrule
RedDots Pre-trained & 47.1 & 44.5 & 49.0 & 39.6 \\
Env-Independent    & 19.9   & 39.9  &  34.6 &  48.9 \\
Env-Dependent       & 13.5 & 17.4 & 21.3 & 22.1 \\
\bottomrule
\end{tabular}
\squeezeup \squeezeup \squeezeup 
\end{table}

To conclude, in this paper, we present the ReMASC data set, which has been built with the goal to advance future research on VCS protection. Compared with previous efforts, the new corpus is much closer to realistic VCS usage scenarios and settings and contains more data variety. Using evaluations with the proposed data set, we find that the performance of the conventional CQCC-GMM model drops significantly when the training and target conditions are mismatched. Defense model trained with data collected in various settings has some, but limited generalization ability to an unseen scenario. Many open research questions can now be studied using this new data set, e.g., can we construct domain-invariant features and models for audio spoofing detection, can we build domain adaptation algorithms to adapt a pre-trained defense model to a new condition, and can we use multi-channel microphone arrays to further improve the defense performance (e.g., using sound directivity features or conducting noise canceling before spoof detection)? The proposed data set can contribute to future research to build stronger and more effective defense models for VCSs.

\newpage

\bibliographystyle{IEEEtran}

\bibliography{mybib}

\end{document}